\documentstyle[12pt]{article}
\newcommand{\be}{\begin{equation}}

\newcommand{\ee}{\end{equation}}
\newcommand{\bea}{\begin{eqnarray}}
\newcommand{\eea}{\end{eqnarray}}
\begin{document}
\title{
\begin{flushright}
{\small SMI-25-98 }
\end{flushright}
\vspace{2cm}
Note on the Massive Rarita-Schwinger Field\\
in the AdS/CFT Correspondence}

\author{
A. S. Koshelev${}^{\S}$ and
O. A. Rytchkov${}^{\star}$,\\
\\${}^{\S}$
{\it Department of Theoretical Physics,  Moscow State University,}\\
{\it Moscow, Russia, 119899}\\
kas@depni.npi.msu.su\\
\\${}^{\star}$
{\it  Steklov Mathematical Institute,}\\
{\it Gubkin str.8, Moscow, Russia, 117966}\\
rychkov@orc.ru
}

\date {$~$}
\maketitle
\begin {abstract}
We consider a massive Rarita-Schwinger field on the Anti-de Sitter
space and solve the corresponding  equations of motion. We show
that appropriate boundary terms calculated on-shell give two-point
correlation functions for spin-3/2 fields of the conformal field
theory on the boundary. The relation between Rarita-Schwinger field
masses and conformal dimensions of corresponding operators is established.
\end{abstract}

\newpage
The correspondence (duality) between supergravity on $d+1$-dimensional
 Anti-de Sitter space and a certain superconformal field theory on
 its $d$-dimensional boundary, proposed in \cite{M} and elaborated in
 \cite{GKP,W} has already undergone a large number of examinations.
 The most important example of this correspondence is provided by
 IIB supergravity on $AdS_5\times S^5$ background which is dual
to ${\cal N}=4$, $D=4$ SYM, appearing on the $AdS_5$ boundary.
Another pair of dual theories involve eleven-dimensional supergravity
compactified on $AdS_7\times S^4$ and (2,0)
superconformal field theory on the $AdS_7$ boundary.
 \par
 According to this correspondence all physical exitations of the bulk
 supergravity can be put in one-to-one correspondence with the
 composite operators of the conformal theory on the boundary and the
 generating functional for these operators is equal to the
 exponentiated classical supergravity action calculated on-shell with
 the given boundary conditions specified by the function $\phi_0$ \be
 \left\langle \exp \left(\int_{\partial AdS_{d+1}}
  d^{d} x {\cal{O}} \phi_0 \right)
 \right\rangle_{CFT}=\exp(-I(\phi)).
 \ee
  Thus in order to find the
 correlation functions of operators in conformal field theory one has
 to solve the classical Dirichlet problem for the corresponding
 supergravity field on the AdS space. It gives the direct way to
 check the correspondence: one should compare the correlation functions
 expected in conformal field theories and those obtained via
 supergravity. In this way two-point functions of
 local operators corresponding to the scalar fields
 \cite{W,AV,MV1,FMMR,AV1}, vector fields \cite{W,FMMR,MV2,CNSS}, 
 spinor fields \cite{HS,MV2}, metric tensor fields \cite{T,AF,MV3}, 
 antisymmetric form fields \cite{AF2,Yi1,Yi2} and massless 
 Rarita-Schwinger fields \cite{Cor,V} were investigated and the 
 agreement with the conformal field results was established. Also 
 various three-point correlation functions 
 for these fields were 
 examined \cite{AV,MV1,FMMR,MV2,GKPF,SSRS}.  
 In all cases the results are in agreement with 
 those expected in the conformal field theory.  
 In this note we consider the massive Rarita-Schwinger field
 and study its two-point correlation function.
 \par
 The general action for the massive Rarita-Schwinger field $\Psi_\mu$
in a $d+1$-dimensional curved space is \cite{Nie}
\be \label{act} S=\int\;
d^{d+1}x\sqrt{g}(\bar\Psi_\mu
\Gamma^{\mu\nu\rho}D_\nu\Psi_\rho-m_1\bar\Psi_\mu\Psi^\mu-m_2\bar\Psi_\mu
\Gamma^{\mu\nu}\Psi_\nu).
\ee
Here $D_\mu$ is a covariant derivative, $\Gamma^{\mu\nu\rho}=
\Gamma^{[\mu}\Gamma^\nu\Gamma^{\rho]}$, $\Gamma^{\mu\nu}=
\Gamma^{[\mu}\Gamma^{\nu]}$, where $\Gamma_\mu$ are curved space
gamma matrices connected with the flat space gamma matrices
$\gamma_a$ via vielbein $e_\mu^a$: $\Gamma_\mu=e_\mu^a\gamma_a$,
$\mu$, $a=0,1,..,d$.
 The flat space
gamma matrices are subjects of the anticommutation
relations $\{\gamma_a,\gamma_b\}=2\delta_{ab}$.  We suppose that the
background is $d+1$-dimensional Euclidean Anti-de Sitter space with
the coordinates $x=(x_0, \vec x)=(x_0, x_i)$, where $x_0>0$,
$i=1,..,d$. The
hypersurface $x_0=0$ will be considered as the AdS boundary.  The
corresponding $AdS$ metric has the form \be
ds^2=\frac{1}{x_0^2}(dx_0^2+dx_idx^i)
\ee
and we can specify the vielbein as
\be
\label{viel}
e_\mu^a=\frac{1}{x_0}\delta_\mu^a.
\ee
Below we will investigate the action (\ref{act}) with rather
general mass parameters $m_1$ and $m_2$. However the most interesting results
are related with the Rarita-Schwinger field on $AdS_5$ appearing in
the compactification of IIB supergravity on $AdS_5\times S^5$.
The corresponding mass spectrum was found in \cite{KRN} and consists
of three sets of states:
\par
 $m_1=-(3/2+k), \quad m_2=0,\quad k=0,1,2,..$
\par
 $m_1=7/2+k, \quad m_2=0,\quad k=0,1,2,..$
\par
 $m_1=0, \quad m_2=-3/2$.
\par

The action (\ref{act}) leads to the following equation of motion
\be
\Gamma^{\mu\nu\rho}D_\nu\Psi_\rho-m_1\Psi^\mu-m_2
\Gamma^{\mu\nu}\Psi_\nu=0,
\ee
which can be rewritten in the equivalent form
\be
\label{eqm}
\Gamma^{\nu}(D_{\nu}\Psi_{\rho}-D_{\rho}\Psi_{\nu})-m_{-}\Psi_{\rho}+
\frac{m_+}{d-1}\Gamma_{\rho}\Gamma^{\nu}\Psi_{\nu}=0,
\ee
where $m_{\pm}=m_1\pm m_2$.
For the future convenience we project the Rarita-Schwinger field
on the vielbein introducing
\be
\psi_a=e_a^\mu\Psi_\mu.
\ee
Substitution of this field and vielbein (\ref{viel}) in Eq.
(\ref{eqm}) is straightforward and gives
$$
x_0\gamma^a\partial_a\psi_b-x_0\partial_b(\gamma^a\psi_a)-\gamma_b
\psi_0-\frac{d}{2}\gamma_0\psi_b-\frac12\gamma_0\gamma_b(\gamma^a
\psi_a)
$$
\be
\label{geneq}
+\frac32\delta_{0b}(\gamma^a\psi_a)-m_{-}\psi_b+\frac{m_{+}}{d-1}
\gamma_b(\gamma^a\psi_a)=0.
\ee
\par
Following Witten \cite{W}, in order to solve these equations
we will find a solution which
depends only on $x_0$ coordinate and has a singularity on the
infinity and then apply the inversion (accompanied by the rotation of
the vielbein), which moves the singularity to the boundary $x_0=0$.
The fields depending only on $x_0$ coordinate
are the subject of the following equations 
\be \label{psi0} 
x_0\gamma^0\partial_0\psi_0-(\frac{d}{2}+1)\gamma_0\psi_0-m_{-}\psi_0=
x_0\partial_0\chi-\frac{m_{+}}{d-1}
\gamma_0\chi-\chi,
\ee
\be
\label{psii}
x_0\gamma^0\partial_0\psi_i-\frac{d}{2}\gamma_0\psi_i-m_{-}\psi_i=
\frac12\gamma_0\gamma_i\chi-\frac{m_{+}}{d-1}
\gamma_i\chi+\gamma_i\psi_0,
\ee
where we denote $\chi=\gamma^a\psi_a$.
It is easy to see that these equations put the constraints
on the fields
\be
\label{restr}
(d-1-2m_{-}\gamma_0)\psi_0=(d-1+2m_2\gamma_0)\chi.
\ee
Taking (\ref{restr}) into account we get the
following solution for the equations (\ref{psi0}), (\ref{psii}) \be
\label{psi0sol}
\psi_0=x_0^{d/2+C}b_0^{+}+x_0^{d/2-C}b_0^{-},
\ee
$$
\psi_i=x_0^{d/2+m_{-}}b_i^{+}+x_0^{d/2-m_{-}}b_i^{-}-
$$
\be
\label{psiisol}
-
\frac{2m_1}{d(d-1-2m_2)}x_0^{d/2-C}\gamma_ib_0^{-}
-\frac{2m_1}{d(d-1+2m_2)}x_0^{d/2+C}\gamma_ib_0^{+},
\ee
where
\be
\label{C}
C=\frac{d(d-1)}{4m_1}+\frac{(m_1-m_2)(m_1+dm_2)}{m_1(d-1)}
\ee
and $b_0^{\pm}$, $b_i^{\pm}$ are constant spinors with a definite
chirality ($\gamma_0b_a^{+}=b_a^{+}$, $\gamma_0b_a^{-}=-b_a^{-}$). 
Also $b_0^{\pm}$, $b_i^{\pm}$ are the subject of
the additional relations \be \gamma^ib_i^{+}=0,\quad
\gamma^ib_i^{-}=0.  \ee
\par
Applying O(d+1,1) transformation of the
  $AdS_{d+1}$ isometry group we can map infinity to an
arbitrary point on the boundary $x_0=0$. It introduces the dependence
on $x_i$ coordinates.  The corresponding transformation is \be
x_\mu\longrightarrow\frac{x_\mu}{x_0^2+\vec x^2}.
\ee
Note that an application of this transformation changes the vielbein
(\ref{viel}), so in order to get a solution in an old local Lorentz
frame we have to transform a vielbein as follows
\be
e_\mu^a\longrightarrow L^a_b e_\mu^b,\quad \mbox{where}\quad
L^a_b=\delta^a_b-\frac{2x^ax_b}{x_0^2+\vec x^2}.
\ee
This inversion of a vielbein is supplemented by an appropriate
transformation of the spinor field \cite{FGP}
\be
\psi_a\longrightarrow L^b_a\frac{x_0\gamma^0+x_i\gamma^i}
{\sqrt{x_0^2+\vec x^2}}\psi_b.
\ee
Note that this transformation changes the sign of gamma-matricies
so  in order to recover the invariance of equations of motion
we have to change all mass signs.
The above procedure, applied to the solution (\ref{psi0sol}),
(\ref{psiisol}) gives us bulk-to-boundary
propagator for the gravitino field. The contraction of this propagator with
arbitrary functions yields

$$ \psi_0=\int\;d^dy
\left(\frac{(x_0\gamma^0+(x-y)_i\gamma^i)x_0^{d/2-C\gamma_0}}{(x_0^2+(\vec
x-\vec y)^2)^{(d+1)/2-C\gamma_0}}(1-\frac{2x_0^2}{x_0^2+(\vec
x-\vec
y)^2})b_0(\vec
y)\right.$$
$$-\frac{2(x_0\gamma^0+(x-y)_i\gamma^i)x_0^{d/2+1-m_{-}\gamma_0}}{(x_0^2+ (\vec
x-\vec y)^2)^{(d+1)/2+1-m_{-}\gamma_0}}(x-y)_ib_i(\vec y)$$
$$
-\frac{4m_1}{d(d-1+2m_2)}\frac{(x_0\gamma^0+(x-y)_i\gamma^i)x_0^{d/2+1+C}}
{(x_0^2+(\vec
x-\vec y)^2)^{(d+1)/2+1+C}}(x-y)_i\gamma_ib_0^{-}(\vec y)$$
\be\label{sol0}\left.-
\frac{4m_1}{d(d-1-2m_2)}\frac{(x_0\gamma^0+(x-y)_i\gamma^i)x_0^{d/2+1-C}}
{(x_0^2+(\vec
x-\vec y)^2)^{(d+1)/2+1-C}}(x-y)_i\gamma_ib_0^{+}(\vec y)\right),
\ee

$$
\psi_i=\int\;d^dy\left(\frac{(x_0\gamma^0+(x-y)_i\gamma^i)x_0^{d/2-m_{-}\gamma_0}}{(x_0^2+
(\vec
x-\vec
y)^2)^{(d+1)/2-m_{-}\gamma_0}}(\delta_{ij}-\frac{2(x-y)_i(x-y)_j}
{x_0^2+(\vec
x-\vec y)^2})b_j(\vec y) \right.
$$
$$-
\frac{2(x_0\gamma^0+(x-y)_i\gamma^i)x_0^{d/2+1-C\gamma_0}}{(x_0^2+
(\vec
x-\vec y)^2)^{(d+1)/2+1-C\gamma_0}}(x-y)_ib_0(\vec y)
$$
$$
+\frac{2m_1}{d(d-1+2m_2)}\frac{(x_0\gamma^0+(x-y)_i\gamma^i)x_0^{d/2+C}}{(x_0^2+
(\vec x-\vec
y)^2)^{(d+1)/2+C}}(\delta_{ij}-\frac{2(x-y)_i(x-y)_j}{x_0^2+ (\vec
x-\vec y)^2})\gamma_jb_0^{-}(\vec y)
$$
\be\label{soli}\left.+
\frac{2m_1}{d(d-1-2m_2)}\frac{(x_0\gamma^0+(x-y)_i\gamma^i)x_0^{d/2-C}}{(x_0^2+
(\vec x-\vec
y)^2)^{(d+1)/2-C}}(\delta_{ij}-\frac{2(x-y)_i(x-y)_j}{x_0^2+ (\vec
x-\vec y)^2})\gamma_jb_0^{+}(\vec y)\right).
\ee
Eq. (\ref{sol0}) and (\ref{soli}) give the general solution
to the equations of motion (\ref{geneq}).
In the same way we construct the solution for the conjugated
spinors $\bar{\psi}_a$, that has the same form except the change
of all mass signs.
\par

Now we have to impose the appropriate boundary conditions on the
infinity and on the hypersurface $x_0=0$.
Namely, in the solution (\ref{sol0}), (\ref{soli}) we keep only the terms,
which vanish as $x_0$ goes to infinity.
Note that the asymptotic behavior of the fields
essentially depends on the values of
the constants $d$, $m_1$ and $m_2$.
Moreover, different terms in (\ref{sol0}), (\ref{soli}) have different
 limits as $x_0$ tends to zero.
In order to be concrete in all formulae below we suppose $m_1$ and
 $m_1-m_2$ to be positive (all other cases could be studied
 in the same way).
 We denote $\psi_a=\psi_a^{'}[b_i^{\pm}]+\psi_a^{''}[b_0^{\pm}]$, where
we collect in $\psi_a^{'}$ all terms depending on $b_i$ and in 
$\psi_a^{''}$ all terms depending on $b_0$. The fields $b_i^{\pm}(\vec
y)$ define the asymptotics of $x_0^{-d/2+m_{-}}\psi_a^{'}[b_i^{\pm}]$.
As in the Dirac case \cite{HS} in order to obtain a square integrable 
asymptotics on the boundary we have to put $b_i^{+}(\vec y)=0$. 
The 
asymptotic behavior of the terms  $x_0^{-d/2+m_{-}}\psi_a^{''}[b_0^{\pm}]$ 
is determined by 
the relation between two constants $m_-$ and $C$. In general, due 
to these relations the functions $b_0$ couldn't be fixed by the 
boundary data, since all the terms 
$x_0^{-d/2+m_{-}}\psi_a^{''}[b_0^{\pm}]$ vanish or go to 
infinity as $x_0$ approaches zero. Thus only $b_i^{-}$ are determined 
by the boundary conditions. 
Note that in the particular case $b_0=0$ we have
the following solution to the Dirichlet problem for the 
Rarita-Schwinger field on the $AdS_{d+1}$ space 
\be\label{sol0_mod} \psi_0(x_0,\vec 
x)=-\int\;d^dy 
\frac{2(-x_0+(x-y)_i\gamma^i)x_0^{d/2+1+m_{-}}}{(x_0^2+ (\vec
x-\vec y)^2)^{(d+1)/2+1+m_{-}}}(x-y)_ib_i^{-}(\vec y),
\ee
\be\label{soli_mod}
\psi_i(x_0,\vec x)=\int\;d^dy\frac{(-x_0+(x-y)_i\gamma^i)x_0^{d/2+m_{-}}}{(x_0^2+(\vec
x-\vec
y)^2)^{(d+1)/2+m_{-}}}\left(\delta_{ij}-\frac{2(x-y)_i(x-y)_j}{x_0^2+(\vec
x-\vec y)^2}\right)b_j^{-}(\vec y).
\ee
\par
The general scheme of AdS/CFT
correspondence \cite{W} prescribes us to calculate
the classical action on-shell with given boundary conditions.  The
action for the Rarita-Schwinger field (\ref{act}) is linear so it
vanishes on-shell. As in the Dirac case \cite{HS} it should be supplied
by a special boundary term, which should be covariant and
quadratic in the Rarita-Schwinger fields \cite{Cor,V}. 
We take such term in the form 
  $I=I_1+I_2$, where
\be\label{bt1} I_1=\mbox{const}\cdot\lim_{\varepsilon\rightarrow
0}\int_{M_d^\varepsilon}\;d^dx \sqrt{g_\varepsilon}\bar\Psi_i(\vec x)
g_\varepsilon^{ij}\Psi_j (\vec x),
\ee
\be\label{bt2}
 I_2=\mbox{const}\cdot\lim_{\varepsilon\rightarrow
0}\int_{M_d^\varepsilon}\;d^dx \sqrt{g_\varepsilon}\bar\Psi_i(\vec x)
\Gamma^i\Gamma^j\Psi_j (\vec x).
\ee
$M_d^\varepsilon$ is the
hypersurface $x_0=\varepsilon$, $\sqrt{g_\varepsilon}$ is a volume element
in the induced metric $g_{ij}$ on this hypersurface. 
 The second structure (\ref{bt2}) doesn't contribute to the
correlation function of operators coupled to $b_i^{-}$ since on-shell 
we have $\gamma^a\psi_a\sim b_0$. In order to calculate $I_1$ we will substitute 
(\ref{soli}) and the corresponding solution for the conjugated spinor 
in (\ref{bt1}) and rewrite it as a sum of the integrals with the 
similar structure. So the integral to be computed is $$ 
G(y_1,y_2,\epsilon)=\int d^dx
\varepsilon^{-d}\left(\delta_{im}-2\frac{(x-y_1)_i(x-y_1)_m}
{\varepsilon^2+(\vec{x}-\vec{y}_1)^2}\right)\times$$ $$
\times\left(\frac{\varepsilon\gamma^0+(x-y_1)_k\gamma^k}
{(\varepsilon^2+(\vec{x}-\vec{y}_1)^2)^{d/2+1/2+\sigma}}\varepsilon^{d/2+\sigma}\right)
\left(\frac{\varepsilon\gamma^0+(x-y_2)_j\gamma^j}
{(\varepsilon^2+(\vec{x}-\vec{y}_2)^2)^{d/2+1/2+\rho}}
\varepsilon^{d/2+\rho}\right)\times$$ \be
\times\left(\delta_{in}-2\frac{(x-y_2)_i(x-y_2)_n}
{\varepsilon^2+(\vec{x}-\vec{y}_2)^2}\right). \label{integral}
\ee
In Appendix we argue that this integral is singular in the limit
$\epsilon\rightarrow 0$
if the difference between $\rho$ and $\sigma$  is not
integer.
Except very special cases of masses the difference
between $C$ and $m_{-}$ is not integer, so  the terms mixing $b_0$
and $b_i$ give a singular contribution in $I_1$.
Therefore if we perform the ''minimal'' renormalization \cite{AV},
i.e. just drop out singular terms, we get
that the generating functional doesn't depend on the
terms containing simultaneously $b_0$ and $b_i$. 
So the correlation functions of operators coupled to $b_i$ are 
determined by the term involving only $b_i$. In order to calculate 
the latter one we will also use the results of Appendix.  In this 
case the hypergeometric function $_3F_2$ in (\ref{genint}) turns to 
$_2F_1$, so we have $$ \int\frac{\varepsilon^{2m_-+1}} 
{(\varepsilon^2+(\vec{x}-\vec{y}_1)^2)^{(d+1)/2+m_-}
(\varepsilon^2+(\vec{x}-\vec{y}_2)^2)^{(d+1)/2+m_-}}
d^dx
$$
$$
=\frac{\mbox{const}}{\varepsilon^{d+2m_-+1}}{}_2F_1
\left(\frac d2+1+2m_-,\frac d2+\frac12+m_-;\frac d2+m_-+1;
-\frac{|\vec{y}_1-\vec{y}_2|^2}{4\varepsilon^2}\right)
$$
\be
=\frac{\mbox{const}}{(4\varepsilon^2+(\vec{y}_1-\vec{y}_2)^2)^{(d+1)/2+m_-}}
{}_2F_1
\left(\frac d2+\frac12+m_-,-m_-;\frac d2+m_-+1;
\frac{|\vec{y}_1-\vec{y}_2|^2}{4\varepsilon^2+|\vec{y}_1-\vec{y}_2|^2}\right)
\label{textint}
\ee
The result
for the integral turns out to be
\be
\label{gen_f}
I_1=\mbox{const}\cdot\int\;d^dy_1d^dy_2\bar
b_i^{+}(\vec y_1)\frac{(y_1-y_2)_k\gamma^k}{|\vec y_1-\vec
y_2|^{d+1+2m_{-}}}(\delta_{ij}-\frac{2(y_1-y_2)_i(y_1-y_2)_j}
{(\vec
y_1-\vec y_2)^2})b_j^{-}(\vec y_2).
\ee
As it was noted in \cite{AV,AV1} the integrals obtained after calculation
the actions on-shell  and taking the limit $\epsilon\rightarrow 0$
are divergent and need the regularization. It was suggested to 
make renormalization or to consider
the integral (\ref{gen_f}) as a value of the distribution 
\be
\frac{(y_1-y_2)_k\gamma^k}{|\vec y_1-\vec
y_2|^{d+1+2m_{-}}}(\delta_{ij}-\frac{2(y_1-y_2)_i(y_1-y_2)_j}
{(\vec
y_1-\vec y_2)^2})
\ee
on a test function. The Fourier transform of this distribution is
\be
F\left[\frac{y_k\gamma^k}{|\vec 
y|^{d+2m_-+1}}(\delta_{ij}-\frac{2y_iy_j} {\vec y^2})\right]=
\mbox{const}\cdot
p^{2m-1}p_k\gamma^k\left(\delta_{ij}+\frac{2}{(d+2m_--3)(d+2m_--1)}
\frac{p_ip_j}{p^2}\right)\label{dist1}
\ee
for $2m_-+1\ne 2k$, where $k=0,1,..$ and $p=|\vec p|$. 
The overall factor in (\ref{dist1}) is not important for us.
Also we take in account that the test functions in (\ref{gen_f})
are the subject of the restriction $\gamma^ib_i=0$. 
When $2m_-+1$ is even, the Fourier transform includes a logarithmic term 
$$ F\left[\frac{y_k\gamma^k}{|\vec 
y|^{d+2m_-+1}}(\delta_{ij}-\frac{2y_iy_j} {\vec y^2})\right]
$$
\be
=\mbox{const}\cdot
p^{2m-1}\log\frac{p}{c}p_k\gamma^k\left(\delta_{ij}+\frac{2}{(d+2m_--3)(d+2m_--1)}
\frac{p_ip_j}{p^2}\right).\label{dist2}
\ee
As it was noted in \cite{AV1} the distributions of this type are not 
conformal invariant.

\par The general scheme prescribes to variate
the unconstrained fields on the boundary, however, we have a
restriction $\gamma_ib_i(\vec y)=0$. It means that we should
introduce a projection operator on this constraint, namely \be
b_i(\vec y)=(\delta_{ij}-\frac1d\gamma_i\gamma_j)\tilde b_j(\vec
y)=\Pi_{ij}\tilde b_j(\vec y), \ee where $\tilde b_j(\vec y)$ is an
unconstrained field. Taking it into account and using (\ref{gen_f})
we obtain
\be\label{corfun} \left\langle \Sigma_{i\alpha}^{+}(\vec y_1)
\Sigma_{j\beta}^{-}(\vec y_2)
\right\rangle=\mbox{const}\cdot\Pi_{ik}\frac{(y_1-y_2)_i
(\gamma^i)_{\alpha\beta}}{|\vec
y_1-\vec
y_2|^{d+1+2m_{-}}}(\delta_{kl}-\frac{2(y_1-y_2)_k(y_1-y_2)_l} {(\vec
y_1-\vec y_2)^2})\Pi_{lj},
 \ee
 $\Sigma_{i\alpha}(\vec y)$ is a conformal field operator which
corresponds to the Rarita-Schwinger field in supergravity.
Note that the second projection operator in (\ref{corfun}) could be
eliminated by the trivial algebraic transformations, so only the
first one is really necessary.
The obtained correlation function coincides with the projected
two-point correlation function expected in the conformal field theory
 for operators with spin-3/2 \cite{cft} and with the conformal
dimension
\be \Delta=\frac{d}{2}+m_{-}.  \ee
For the
massless case $m_1=0$ these results coincide with those obtained in
\cite{Cor,V}.  \par

In this note we have considered the action
for the massive Rarita-Schwinger field in the Anti-de Sitter space
and solve the corresponding equations of motion. According to the
prescriptions of AdS/CFT correspondence we have demanded the
appropriate boundary behavior of the field and using the special
boundary terms we have obtained the two-point correlation function
for the conformal field theory operators with spin-3/2.
In our consideration we didn't specify the AdS dimension $d$ and the
parameters $m_1$ and $m_2$ of the Rarita-Schwinger field.
However we implied that these parameters are subject of the special
relations:  we supposed that the difference between $C$ and $m_{-}$
is not integer. It is easy to see that for the Rarita-Schwinger field
masses in IIB supergravity on $AdS_5\times S^5$ \cite{KRN} and in
eleven-dimensional supergravity on $AdS_7\times S^4$ \cite{PvN} the
above condition is satisfied.

$$~$$

{\bf ACKNOWLEDGEMENTS}
$$~$$
We are grateful to I. Ya. Aref'eva for suggesting this problem
and permanent interest to our work.
Also we are grateful to G. E. Arutyunov for
useful and stimulating discussions.  A.K. and O.R.  are supported in
part by RFFI grant 96-01-00608. O.R. is supported in part by the
INTAS Grant 96-0457
within the research program of the International Center for
Fundamental Physics in Moscow.

\section*{Appendix}

The purpose of this Appendix is to give an explicit calculation of
the general integral (\ref{integral}).
Performing multiplication, one can get
$$
G(y_1,y_2,\epsilon)=\int
d^dx\varepsilon^{\rho+\sigma} 
\frac{\varepsilon^2+(x-y_1)_k\gamma^k(x-y_2)_j\gamma^j-\varepsilon(y_1-y_2)_k\gamma^k\gamma_0}
{(\varepsilon^2+(\vec{x}-\vec{y}_1)^2)^{d/2+1/2+\sigma}
(\varepsilon^2+(\vec{x}-\vec{y}_2)^2)^{d/2+1/2+\rho}}
\times$$ $$
\times\left(\delta_{mn}+4\frac{(x-y_1)_i(x-y_2)^i(x-y_1)_m(x-y_2)_n}
{(\varepsilon^2+(\vec{x}-\vec{y}_1)^2)
(\varepsilon^2+(\vec{x}-\vec{y}_2)^2)}\right.$$ \be\left.
-2\frac{(x-y_1)_m(x-y_1)_n}
{\varepsilon^2+(\vec{x}-\vec{y}_1)^2}-2\frac{(x-y_2)_m(x-y_2)_n}
{\varepsilon^2+(\vec{x}-\vec{y}_2)^2}\right).
\label{fullintegral}
\ee

We will consider (\ref{fullintegral}) as a sum of different terms
with the similar structure, it will unify our investigation.
Also the succeeding analysis will be simplified if we rewrite each term
applying the Fourier transformation.
We use the following Fourier transformation formula:
\be
\label{fur1}
2^{\alpha}\Gamma\left(\frac d2 +\alpha\right)\frac{\varepsilon^{\alpha}}{(\varepsilon^2+(\vec{x}-\vec{y})^2)^{d/2+\alpha}}=
\mbox{const}\cdot
\int
d^dpe^{i\vec{p}(\vec{x}-\vec{y})}p^{\alpha}K_{\alpha}(\varepsilon
p)
\ee
where $K_{\alpha}$ is a modified Bessel function with index $\alpha$ and $p=|\vec{p}|$.
The contraction of two propagators in (\ref{fullintegral}) turns into the
product of their Fourier images. The remaining integral could be found
explicitly in \cite{Prud}, so we get
\bea
&&\int e^{-i\vec{p}(\vec{y_1}-\vec{y_2})}
p^{\mu+\nu}K_{\mu}(\varepsilon p)K_{\nu}(\varepsilon p)
d^dp=\nonumber\\
&&=\frac{\mbox{const}}{\varepsilon^{d+\mu+\nu}}
\frac{2^{d/2+\mu+\nu-2}\Gamma\left(d/2+\mu+\nu\right)
      \Gamma\left(d/2+\nu\right)
      \Gamma\left(d/2+\mu\right)}
     {\Gamma\left(d+\mu+\nu\right)}
      \times\nonumber\\
&&_3F_2\left(d/2+\mu+\nu,
           d/2+\nu,
           d/2+\mu;
           \frac{d+\mu+\nu}2,
           \frac{d+\mu+\nu+1}2;
           -\frac{|\vec y_1-\vec y_2|^2}{4\varepsilon^2}\right).
\label{genint}
\eea
All factors like $p_m$ appearing after differentiation
of (\ref{fur1})
could be obtained by applying the operator $i\frac{\partial}{\partial
y_m}$ to (\ref{genint}).  The circle of convergence for the
hypergeometric function $_3F_2(z)$ is $|z|<1$.  So, in order to find
the limit $\varepsilon\rightarrow 0$ we have to apply the formula of
analytic continuation, known in the theory of the hypergeometric
functions $$ _{3}F_2(a_1,a_2,a_3;b_1,b_2;z)=
\frac{\Gamma(b_1)\Gamma(b_2)}{\Gamma(a_1)\Gamma(a_2)\Gamma(a_{3})}
\left[\Gamma(a_1)(-z)^{-a_1}\frac{\Gamma(a_2-a_1)\Gamma(a_3-a_1)}
{\Gamma(b_1-a_1)\Gamma(b_2-a_1)}\times\right.$$ $$
{}_3F_2(1+a_1-b_1,1+a_1-b_2,a_1;1+a_1-a_2,1+a_1-a_3;1/z)$$
$$+\Gamma(a_2)(-z)^{-a_2}\frac{\Gamma(a_1-a_2)\Gamma(a_3-a_2)}
{\Gamma(b_1-a_2)\Gamma(b_2-a_2)}\times$$ $$
\times{}_3F_2(1+a_2-b_1,1+a_2-b_2,a_2;1+a_2-a_1,1+a_2-a_3;1/z)$$ $$
+\Gamma(a_3)(-z)^{-a_3}\frac{\Gamma(a_1-a_3)\Gamma(a_2-a_3)}
{\Gamma(b_1-a_3)\Gamma(b_2-a_3)}\times$$ \be
\left.\times\frac{}{}{}_3F_2(1+a_3-b_1,1+a_3-b_2,a_3;1+a_3-a_1,1+a_3-a_2;1/z)\right].
\label{invert}
\ee
Taking into account, that $F(0)=1$ we find three terms with the different
behavior as $\varepsilon$ tends to zero. In our concrete case of the integral
(\ref{fullintegral}) we have $\mu=\sigma+1/2-n$, $\nu=\rho+1/2-m$,
where $m$, $n$ are integers, which are different for each term in
 (\ref{fullintegral}). So we find that three mentioned types of
behavior as $\varepsilon\rightarrow 0$
are
\be
\varepsilon^{\sigma+\rho+d+1-n-m},\qquad
\varepsilon^{\sigma-\rho-n+m},\qquad \varepsilon^{\rho-\sigma+n-m}.
\ee
Note that the powers in the second and the third types are inverse
to each other, so if the difference between $\rho$ and $\sigma$  is not
integer we have a singular contribution.

\end{document}